\documentclass[final,5p,times,twocolumn]{elsarticle}

\newcommand{\ba}{\begin{eqnarray}}
\newcommand{\ea}{\end{eqnarray}}
\newcommand{\be}{\begin{equation}}
\newcommand{\ee}{\end{equation}}
\newcommand{\eps}{\epsilon}

\usepackage{amssymb}

\usepackage{lineno}

\usepackage[T1]{fontenc}
\journal{Physics Letters A}
\usepackage{color}
\begin{document}

\begin{frontmatter}

\title{Modulational instability in wind-forced waves}

\author[label1]{Maura Brunetti}
\author[label2]{J\'er\^ome Kasparian}

\address[label1]{GAP-Climate and Institute for Environmental Sciences,
University of Geneva, Route de Drize 7, 1227 Carouge, 
Switzerland}
\address[label2]{GAP-Nonlinear, University of Geneva, 
Chemin de Pinchat 22, 1227 Carouge, Switzerland}

\begin{abstract}

We consider the wind-forced nonlinear Schr\"odinger (NLS) 
equation obtained in the potential flow framework 
when the Miles growth rate is of the order of the wave steepness. In this case, the form of the wind-forcing terms  
gives rise to the enhancement of the modulational instability and to a band of positive gain with infinite width. This regime is characterised by the fact that the ratio between wave momentum and norm is not a constant of motion, in contrast to what happens in the standard case where the Miles growth rate is of the order of the 
steepness squared.  
\end{abstract}

\begin{keyword}
Modulational instability  \sep Wind forcing \sep Water waves  \sep Rogue waves

\end{keyword}

\end{frontmatter}


\section{Introduction}
\label{intro}

The modulational instability (known as Benjamin-Feir instability in the context of 
fluid dynamics~\cite{BF1967,Zakharov1968}) is 
ubiquitous in physics, it occurs 
in nonlinear waves  within numerous 
physical situations (water waves, plasma waves, laser beams, electromagnetic transmission lines,...)~\cite{2009PhyD..238..540Z,2011PhRvA..83a3830B} 
and it is one of the possible mechanisms of catastrophic growth and  
generation of rogue waves in the ocean~\cite{2009PhLA..373..675A}. 

The stability properties of the wavetrains 
rely on the form of the damping/pumping terms in the governing equations which, in the context of water waves, depend on the wind providing energy to the system~\cite{Fabrikant1980,Bridges2007,Kharif2010}. 
Modeling the effects of wind on ocean waves is a very complex task due to turbulence in 
both the atmospheric and the oceanic boundary layers, and nonlinearities in the propagation 
of the gravity waves at the interface. The problem has been simplified by 
assuming quasi-laminar airflows~\cite{blennerhassett1980} through the Miles mechanism~\cite{Miles1957}, 
quasi-linear theory in wind-wave generation 
(the Janssen mechanism~\cite{Janssen1991}) 
and different approximations in the wave dynamics 
(i.e. in the Navier-Stokes equations or the Euler equations) to obtain mathematical models for the propagation of surface gravity waves which can be handled analytically. 
The wind can induce either damping or forcing terms in the resulting equations~\cite{Leblanc2007,Kharif2010} depending on its speed and direction relative to the wave propagation.  
Many experiments have been performed to investigate how surface 
waves and modulational instability are affected by wind and dissipation~\cite{1986JFM...162..237B,1999JFM...401...55W,2005JFM...539..229S,2013JFM...722....5G,2013PhFl...25j1704C}, sometimes with contrasting results regarding 
in particular the 
values of the damping rates induced by winds blowing slower or opposite to the wave velocity~\cite{1983Donelan,1985JFM...151..427Y,2003JFM...487..345P}.  
 
The effect of wind can be modelled  in the framework of the Miles 
mechanism~\cite{Miles1957} and the potential flow approximation~\cite{2008Dias} for deep-water waves.
The 
growth rate $\Gamma_M/f$ of the wave energy (normalised with respect to the frequency of the carrier wave)  
is most often taken of the same order as the dissipation, hence at the
$\Gamma_M/f = O(\eps^2)$, and the resulting  envelope equation at third-order in the wave steepness $\eps$ is given 
by a wind-forced nonlinear Schr\"odinger (NLS)  equation~\cite{Leblanc2007,Kharif2010,OnoratoProment2012} of the form
\be
i\frac{\partial a}{\partial t} - \beta_1\frac{\partial^2 a}{\partial x^2} - M |a|^2 a = i \left(\frac{\Gamma_M}{2} - 2\nu k^2 \right)a
\label{dampedForced}
\ee 
where $\beta_1 = -(dc_g/dk)/2 = \omega/(8k^2)$,  $M = \omega k^2/2$, 
and $\nu$ is the kinematic viscosity.
 
Recently we have derived the wind-forced NLS for stronger wind forcing, with a 
growth rate $\Gamma_M/f$ of the wave energy   
of the same order as the 
steepness~\cite{2014PhLA..378.1025B}, $\Gamma_M/f = O(\eps)$. In this case, the envelope 
equation obtained by 
the multiple-scale perturbation method at third-order in  $\eps$ reads
\be
i \frac{\partial a}{\partial t} -\beta_1  \frac{\partial^2 a}{\partial x^2} -M a |a|^2 =  \left( \beta_2
\frac{\partial }{\partial x}+\beta_3  -2i\nu k^2 \right)a
\label{NLSwind} 
\ee
where  
$\beta_2 = 3\Gamma_M/(4k)$ and 
$\beta_3 =  \Gamma_M^2 / (8\omega)$.
As compared to eq.~(\ref{dampedForced}), the latter equation contains two additional forcing terms, namely the terms proportional to $\beta_2$ and $\beta_3$.

In this Letter, we investigate the effects of the wind-forcing terms in eq.~(\ref{NLSwind}) 
on the modulational instability (section 2) and compare it to the well-known case 
described by eq.~(\ref{dampedForced}) 
for reference. We show that considering the wave-energy growth rate at the first order in 
steepness results in widely extending the spectral range of the modulational instability gain.
Besides, we show (section 3) that the way wind-forcing is considered affects the ratio of the momentum to the norm of the pulse, that is conserved only if the growth rate is limited to the second order in steepness.
We compare this finding with recent
sets of experiments where 
either the  carrier wave amplitude or the 
initial perturbation amplitudes are sufficiently large and the modulational instability is 
enhanced~\cite{2005JFM...539..229S}, suggesting the physical relevance of considering the model 
given by eq.~(\ref{NLSwind}). We discuss the main results in section 4 and we draw the conclusions in section 5.    

\section{Modulational instability} 

Benjamin and Feir~\cite{BF1967} showed that inviscid deep-water wavetrains are unstable to 
small perturbations of other waves travelling in the same direction with frequencies within 
the band of positive gain. We compare here the modulation instability  
when wind forcing terms are included in the envelope 
equations in two different regimes: low Miles growth rates 
$\Gamma_M/f = O(\epsilon^2)$ (that is the well-known standard case that we develop for reference) and 
high Miles growth rates $\Gamma_M/f = O(\epsilon)$.

\subsection{Low growth rates}

We review here for reference the standard case where the envelope equation is given by eq.~(\ref{dampedForced}). This will be useful to set-up the formalism and to compare with results 
obtained when considering growth rates at the first order in steepness. 

By defining $\tau = \omega t$, $\xi = 2k x$, $\Gamma = \Gamma_M/(2\omega)$, 
$\delta = 2\nu k^2/\omega$, $K = \Gamma-\delta$, and $A =  k a/\sqrt{2}$, 
the equation (\ref{dampedForced}) reduces to~\cite{Kharif2010}
\be
i A_\tau -  \frac{1}{2} A_{\xi\xi}  -A |A|^2 = iK A
\label{NLSwind1} 
\ee
The factor $K$ on the right-hand side can be positive, null or negative depending on the relative 
importance of the viscosity term $\delta$ with respect to the wind-forcing term $\Gamma$.  
The Stokes-like wave, which is a solution of eq.~(\ref{NLSwind1}) 
independent on $\xi$, is given by 
\be
A_S(\tau) = A_0\, e^{K\tau} e^{-i b(\tau) } \, , \qquad b(\tau) = \frac{|A_0|^2}{2 K}
( e^{2K \tau}-1) 
\ee
Note that for $K = 0$, we get $b(\tau) = |A_0|^2 \tau$, which is valid in the inviscid 
case. Following previous studies~\cite{2005JFM...539..229S,Leblanc2007,Kharif2010}, the Stokes-like wave is perturbed as follows
\be
A(\xi,\tau) = A_S(\tau) [1+\delta_0\, \zeta(\xi,\tau)]
\label{perturbStokes}
\ee
with $\delta_0$ infinitesimal and $\zeta(\xi,\tau) = M(\xi,\tau)+iN(\xi,\tau)$. 
Substituting into eq.~(\ref{NLSwind1}) gives the following system of equations
\ba
M_\tau -\frac{1}{2} N_{\xi\xi} &=& 0 \\
N_\tau  +\frac{1}{2} M_{\xi\xi} + 2|A_S|^2 M &=& 0
\ea
By choosing perturbations of the form 
\ba
M(\xi,\tau) &=& \Re \{ M_0(\tau)\, e^{i\ell \xi} \} \label{mi1} \\
N(\xi,\tau) &=& \Re \{ N_0(\tau)\, e^{i \ell\xi}\}
\label{mi2}
\ea
where $\ell$ is the modulational wavenumber, the previous system becomes
\ba
\frac{d M_0}{d\tau} +\frac{\ell^2}{2} N_0 &=& 0 \\
\frac{d N_0}{d\tau}  - \left(\frac{\ell^2}{2}   - 2|A_S|^2\right) M_0 
&=& 0
\ea
which corresponds to the following 
equation~\cite{2005JFM...539..229S,Leblanc2007,Kharif2010}
\be
\frac{d^2 M_0}{d\tau^2} +\frac{\ell^2}{2} 
\left(\frac{\ell^2}{2}  -  2|A_0|^2\, e^{2K \tau}\right) M_0 = 0
\label{standardMI}
\ee
In the case $K=0$, this differential equation has constant coefficients and by setting $M_0(\tau) = \tilde M\, e^{-i\Omega \tau}$, one gets the  dispersion relation~\cite{2005JFM...539..229S}
\be
\Omega = \pm \frac{\ell}{\sqrt{2}} 
\sqrt{\frac{\ell^2}{2}  -  2\, |A_0|^2}
\label{DispRelCase1}
\ee 
In the case $K\not = 0$, eq.~(\ref{standardMI}) is a Sturm-Liouville problem~\cite{2005JFM...539..229S} which must be analysed as in~\cite{2005JFM...539..229S,Kharif2010}. The presence of oscillatory or exponentially growing solutions depends on the sign of the factor $(\ell^2/2  -  2|A_0|^2\, e^{2K \tau})$ in eq.~(\ref{standardMI}).  
Growing perturbations of the Stokes-like solution appear in a limited range of modulational 
wavenumbers~\cite{2005JFM...539..229S}
\be
\ell^2  < 4 |A_0|^2 \, e^{2 K\tau}
\ee 
The stability range expands (contracts) with time in the presence of pumping $K= \Gamma-\delta>0$ (damping $\Gamma<
\delta$), but the Benjamin-Feir instability gain is independent from the pumping/damping 
term~\cite{2005JFM...539..229S,Bridges2007}. In other words, the dependence on $\Gamma$ is only 
within the exponential term which 
appears in the Stokes wave amplitude $|A_S|$ and determines expansion or contraction depending on the sign of 
$K$. The range where modulational wavenumbers become unstable is shown in Fig.~1,
dashed line,  for $|A_S| = 0.1$. 
The maximum growth rate occurs at $\ell^* = \pm \sqrt{2}\, |A_S|$  
(see vertical dash-dotted lines in Fig.~1)   
\be
\Omega_{I}(\ell = \ell^*) = |A_0|^2\,  e^{2K \tau} = |A_S|^2
\label{OmegaMax1}
\ee   

\begin{figure}
\centering
\includegraphics[width=0.5\textwidth]{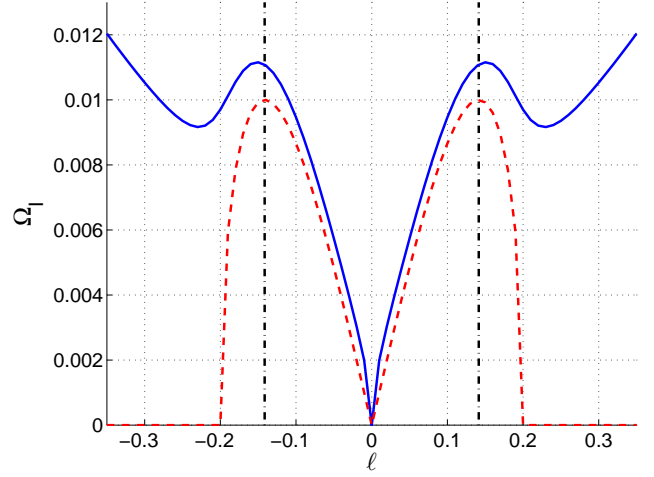}
\caption{Band of positive gain for the modulational wavenumbers $\ell$ with $|A_S| = 0.1$: low Miles growth-rates
(dashed red line) and high Miles growth-rates (solid blue line) with $\Gamma_M/f = \eps  = \sqrt{2}\, |A_S|$. Dash-dotted vertical lines correspond to 
$ \ell^* = \pm \sqrt{2}\, |A_S|$.}
\label{fig:1}
\end{figure}

\subsection{High growth rates}

Here we conduct a similar procedure in the case of the 
envelope equation (\ref{NLSwind}) obtained from the full nonlinear gravity-wave equations when 
the Miles growth rate is $\Gamma_M/f = O(\epsilon)$~\cite{2014PhLA..378.1025B}. 
By defining as before $\tau = \omega t$, $\xi = 2k x$, $\Gamma = \Gamma_M/(2\omega)$, 
$\delta = 2\nu k^2/\omega$, and $A = ka /\sqrt{2}$, 
this equation reduces to
\be
i A_\tau -  \frac{1}{2} A_{\xi\xi}  
-A |A|^2 =  3\Gamma\,  A_\xi +\frac{1}{2} \Gamma^2 A -i\delta A
\label{NLSwind2} 
\ee
The Stokes-like wave is given by: 
\be
A_S(\tau) = A_0\, e^{-\delta \tau}e^{-ib(\tau) }\, , \, \, b(\tau) = \frac{ \Gamma^2/2 + |A_0|^2 }{-2\delta}(e^{-2\delta \tau}-1) 
\ee
Now if we perturb the Stokes-like solution as stated previously (eq.~(\ref{perturbStokes})) and we 
substitute the perturbed wave into eq.~(\ref{NLSwind2}), we obtain the following system of equations
\ba
M_\tau - 3\Gamma N_\xi -\frac{1}{2} N_{\xi\xi} &=& 0 \\
N_\tau + 3\Gamma M_\xi +\frac{1}{2} M_{\xi\xi} + 2|A_S|^2 M &=& 0
\ea
By choosing perturbations of the form given by eqs.~(\ref{mi1})-(\ref{mi2}), the system becomes
\ba
\frac{d M_0}{d\tau} +\left( \frac{\ell^2}{2}- 3i\,\Gamma \ell \right) N_0 &=& 0 \\
\frac{d N_0}{d\tau}  - \left(\frac{\ell^2}{2}  - 3i\ell \Gamma - 2|A_S|^2\right) M_0 
&=& 0
\ea
which corresponds to the following equation
\be
\frac{d^2 M_0}{d\tau^2} +\left(\frac{\ell^2}{2}-3i\ell \Gamma\right) 
\left(\frac{\ell^2}{2}  - 3i\ell \Gamma - 2|A_0|^2\, e^{-2\delta \tau}\right) M_0 = 0
\label{diffEq}
\ee
This equation must be compared to eq.~(\ref{standardMI}): in the present case there are two additional imaginary terms within the parenthesis. 
If we neglect viscosity (by setting $\delta = 0$), this differential equation has constant 
coefficients\footnote{When viscosity cannot be neglected, eq.~(\ref{diffEq}) is a Sturm-Liouville problem and must be analysed with the same approach used in \cite{2005JFM...539..229S,Kharif2010}.} and by setting $M_0(\tau) = \tilde M\, e^{-i\Omega \tau}$, we get the dispersion relation
\ba
\Omega &=& \pm \sqrt{\left(\frac{\ell^2}{2}-3i\ell\Gamma\right) 
\left(\frac{\ell^2}{2}  - 3i\ell \Gamma - 2|A_S|^2 \right)} \label{dispRelWind}\\
&=& \pm \sqrt{\Omega_1+ i\Omega_2} = \pm (\Omega_R + i \Omega_I)
\label{dispRelWind2}
\ea
where
\ba
\Omega_1 &=& \frac{\ell^2}{2}\left(\frac{\ell^2}{2}-2|A_S|^2-18\Gamma^2\right) \\
\Omega_2 &=& -3\ell \,\Gamma \left(\ell^2 - 2|A_S|^2\right) 
\ea
and the real and imaginary part are
\ba
\Omega_R &=& \sqrt{\frac{\Omega_1 + \sqrt{\Omega_1^2 + \Omega_2^2}}{2}} \\
\Omega_I &=&  \sqrt{\frac{-\Omega_1 + \sqrt{\Omega_1^2 + \Omega_2^2}}{2}}
\ea
In contrast with the standard case, $\Omega_I$ is defined for all the modulational wavenumbers, 
as shown in Fig.~1 (solid line). Furthermore, 
for  $\ell = \ell^* = \pm \sqrt{2}\, |A_S|$, the growth rate is given by  
\be
\Omega_I(\ell=\ell^*) = |A_S|^2 \sqrt{1+ \frac{18\,\Gamma^2}{|A_S|^2}} 
\label{OmegaMax2}
\ee
Comparing this expression with the corresponding value in eq.~(\ref{OmegaMax1}), it is obvious that the form of wind forcing on the right-hand side of eq.~(\ref{NLSwind}) 
 enhances not only the width of the modulational instability band, but also the gain of the Benjamin-Feir instability at each frequency, as can be 
seen in Fig.~1, solid line. The enhancement is directly related to the forcing factor $\Gamma$ in 
eq.~(\ref{dispRelWind}) which, contrary to the second-order case, does not appear in the Stokes wave amplitude 
$|A_S|$. More surprising, the enhancement occurs regardless of the sign of $\Gamma$, i.e. for both wind forcing and damping, as can be seen from eq.~(\ref{OmegaMax2}). 

This can be understood by considering that the first term on the right-hand side of eq.~(\ref{NLSwind2}) plays the role of a modification of  the wave group velocity under the effect of wind~\cite{Fabrikant1980}, i.e. a modification of the coefficient of the second term on the left-hand side of eq.~(\ref{NLSwind2}). It therefore contributes to the phase matching necessary to trigger efficient modulational instability. 
More specifically, since $\partial_\xi$, $A$ and $\Gamma$ are $O(\eps)$-terms, at third-order in $\eps$ (which is the order in the multiple-scale 
approach where the NLS equation is obtained) we can replace $\Gamma A_{\xi\xi} \sim ik\, \Gamma A_\xi$ 
and thus the two considered terms in eq.~(\ref{NLSwind2}) can be rewritten as  
\be
-\frac{1}{2} A_{\xi\xi} - 3\Gamma A_\xi \sim - \left(\frac{1}{2} + \frac{3\Gamma}{i k} \right)A_{\xi\xi} 
\ee
so that the NLS equation takes the form of eq.~(1) in Ref.~\cite{Bridges2007} with $a\not = 0$, which 
indeed corresponds to the regime 
where negative energy modes\footnote{Energy is relative to that of the carrier wave 
$E_{\rm Stokes}$, so that `negative energy' means that $E-E_{\rm Stokes }<0$.} can be destabilised and Benjamin-Feir instability enhanced.  

As can be seen from Fig.~1, the gain $\Omega_{I}$ has a local maximum  which occurs for modulational wavenumber slightly larger than the standard case\footnote{The analytic expression can be easily found using 
symbolic programs like Mathematica, but it is very long and not particularly illuminating.}, $|\ell_{max}| \gtrsim \sqrt{2}\, |A_S|$. For large modulational wavenumbers, the dependence of the imaginary part 
$\Omega_I$ becomes asymptotically linear, $\Omega_I \sim 3 |\ell|\, \Gamma$.  
Moreover, the dependence of $\Omega_I$ on $\Gamma_M/f$ is   shown in Fig.~2, where we can see that the local maximum disappears at growth rates of the order of $\Gamma_M/f \sim 
1.5\,\epsilon$.   

The final spectrum generated by the modulational instability depends on both the seed of the initial spectrum, and the gain accumulated  over propagation for each spectral component. In the case of the modulational-instability gain as generated by low growth rates, $\Gamma_M/f = O(\epsilon^2)$, the bandwidth is intrinsically limited but expands in time (for $K>0$) to reach high modulational wavenumbers~\cite{Leblanc2007}. 
Conversely, for high growth rates, $\Gamma_M/f = O(\epsilon)$, 
the modulational-instability gain band has infinite width from the initial time, thus inducing broadening of the initial spectrum and development of turbulence in the presence of either damping or pumping force ($\Gamma <$ or $> 0$). 
Numerical simulations are required to understand the role of the high modulational wavenumbers as a function of the initial spectrum under different parameterisation of the wind forcing, and this will be the subject of a forthcoming paper.   

\begin{figure}
\centering
\includegraphics[width=0.5\textwidth]{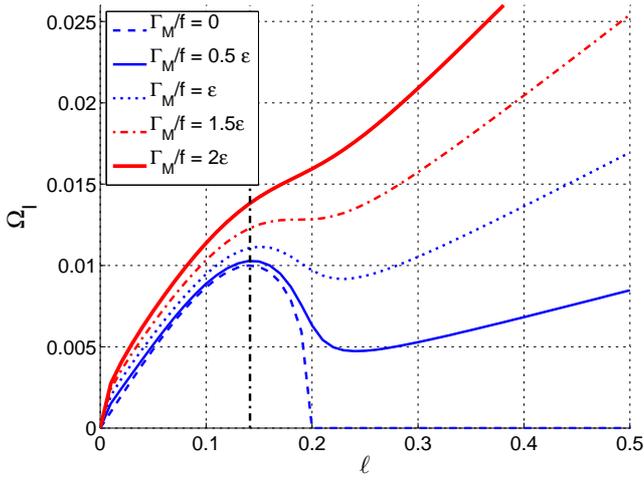}
\caption{Band of positive gain for the modulational wavenumbers $\ell$ with $|A_S| = 0.1$ for different 
values of the Miles growth-rate $\Gamma_M/f$.}
\label{fig:2}
\end{figure}

\section{Effect of the wind-forcing term on the momentum to norm ratio}

The form of the wind forcing as proposed in eq.~(\ref{NLSwind})  qualitatively affects 
not only the modulational instability, but also the conservation of the ratio of the wave momentum to its norm, that are respectively defined as~\cite{2005JFM...539..229S}
\be
P = -i \int A_x A^* dx
\ee
and
\be
N = \int |A|^2 dx    
\ee
Adding a forcing term of the form $i K A$ as in eq.~(\ref{NLSwind1}) destroys the conservation of 
$N$ and $P$, which then evolve in time as follows: 
\ba
N(t) &=& N_0\, e^{2K t} \\
P(t) &=& P_0\, e^{2K t} 
\ea 
where $K = \Gamma - \delta$.
Note that the ratio between $P$ and $N$ is constant, $P/N= P_0/N_0$~\cite{2005JFM...539..229S}. 

In contrast, the terms proportional to $\Gamma$ and $\delta$ in eq.~(\ref{NLSwind2}) modify the 
temporal evolution of 
$N$ and $P$ as follows: 
\ba
\frac{dN}{dt} &=& 6\Gamma P - 2 \delta N \\
\frac{dP}{dt} &=& 6\Gamma \int |A_x|^2 dx - 2\delta P 
\ea 

The ratio between $P$ and $N$ satisfies the following equation
\be
\frac{d}{dt}\left(\frac{P}{N}\right) = \frac{6\Gamma}{N}\left(\int |A_x|^2 dx - \frac{P^2}{N}\right)
\ee 
Since $(-i\int A_x A^* dx)^2 \not = (\int |A_x|^2 dx) (\int |A|^2 dx)$, the right-hand side is in general not zero and thus the ratio $P/N$ is not constant in time for forcing with $\Gamma_M/f = O(\eps)$. 
Note that the latter relation does not depend on the viscosity $\delta$ since the corresponding terms canceled out.  Thus, we have found that the evolution of the ratio $P/N$ characterises the two different regimes,  $\Gamma_M/f = O(\eps^2)$ and $\Gamma_M/f = O(\eps)$, and it can be used 
in experiments to check different parameterisations of the forcing terms in the envelope equation. 

\section{Discussion}

The above calculations show that considering the wind forcing with wave-energy growth rate 
$\Gamma_M/f$ at the first order in the steepness $\eps$ substantially affects the model outcome. Although the damping of ocean waves in an adverse wind is not adequately 
modeled by the Miles mechanism~\cite[Chap.~3]{JanssenBook}, this damping is important and several investigations have found damping rates comparable to the corresponding growth 
rates~\cite{1983Donelan,2003JFM...487..345P}.  If this is confirmed, the above work can be also applied if $\Gamma_M$ stands for a damping term: only its sign needs to be inverted.

Our finding can therefore be connected with experiments showing that, in the presence of dissipation, wavetrains with moderate carrier-wave amplitudes conserved the ratio $P/N$, while those 
with large carrier-wave or perturbation amplitudes led to not-conserved $P/N$ and to enhanced modulational instability~\cite{2005JFM...539..229S}. These results cannot be described by damping modeled as in eq.~(\ref{dampedForced}), while they can be explained by a model, such as eq.~(\ref{NLSwind}), where wave-energy damping rates are assumed of the order of the steepness, since it formally predicts that $P/N$ is not 
a constant of motion and broadening is enhanced via stronger modulational instability .   
New experiments are required to test in detail this hypothesis. 

\section{Conclusions} 

The modulational instability is a fundamental mechanism for nonlinear 
exchanges of energy between carrier and sideband waves. It is ubiquitous in 
physics~\cite{2009PhyD..238..540Z} and it is one of the 
mechanism of rogue-waves formation in deep-water~\cite{2009PhLA..373..675A}. Since the wind is the energy source in surface wave propagation, it is expected that accurate 
modeling of the wind is critical for understanding rogue wave formation.

We have investigated how different forcing/damping terms, due to the wind action, affect the band of positive gain of the modulational instability.  
In particular, we have considered the recently proposed model of envelope 
waves~\cite{2014PhLA..378.1025B} which 
is obtained from the full nonlinear gravity waves equations  by assuming potential flow and 
the Miles mechanism for the growth of ocean waves under wind action. Modelling the wind forcing (or, equivalently, the damping) with rates at the first order in wave steepness qualitatively affects the modulational instability as well as the conservation of the momentum to norm ratio of the wave, as compared to weaker rates.

We find that the proposed parameterisation of the wind forcing gives rise to the enhancement 
of the modulational instability, as shown in Fig.~1.
The enhanced modulational instability is attributed to the fact that the forcing term in the wind-forced envelope equation~(\ref{NLSwind}) is equivalent to a correction of the wave group velocity under the effect of wind~\cite{Fabrikant1980,Bridges2007}, hence allowing phase matching that would be inaccessible without it. 
Thus the proposed parameterisation corresponds to the $a$-term in the NLS model for dissipatively perturbed Stokes waves in deep water 
considered in~Ref.~\cite{Bridges2007}, which indeed leads to the enhancement of the modulational instability.    

Furthermore, the transition to a larger modulational instability as well as a loss in the $P/N$ ratio conservation for larger growth rates offers an interpretation to previously published experimental results~\cite{2005JFM...539..229S} showing such transition for increased carrier-wave or perturbation amplitudes. 
It therefore illustrates the need to consider all mechanisms of energy exchange with the wave, including dissipation as well as wind-forcing, at their right order to avoid underestimating them.

In summary, we have found a form of wind forcing which enhances the modulational instability and gives rise to a ratio between momentum and norm which is not conserved in time. Tank and numerical 
experiments are needed to confirm if these effects are physically realisable. The 
enhancement of the modulational instability on broad and narrow-banded spectra must be 
analysed through numerical simulations to understand in particular the role of high-frequencies sidebands on the envelope evolution. 

\medskip
\noindent
{\small  We thank Prof. Fr\'ed\'eric Dias for useful discussions.} 

\bibliographystyle{elsarticle-num}
\bibliography{waves}

\begin{thebibliography}{10}
\expandafter\ifx\csname url\endcsname\relax
  \def\url#1{\texttt{#1}}\fi
\expandafter\ifx\csname urlprefix\endcsname\relax\def\urlprefix{URL }\fi
\expandafter\ifx\csname href\endcsname\relax
  \def\href#1#2{#2} \def\path#1{#1}\fi

\bibitem{BF1967}
T.~{Benjamin}, J.~{Feir}, {The disintegration of wavetrains in deep water. Part
  1}, Journal of Fluid Mechanics 27 (1967) 417--430.

\bibitem{Zakharov1968}
V.~E. {Zakharov}, {Stability of periodic waves of finite amplitude on the
  surface of a deep fluid}, Journal of Applied Mechanics and Technical Physics
  9 (1968) 190--194.

\bibitem{2009PhyD..238..540Z}
V.~E. {Zakharov}, L.~A. {Ostrovsky}, {Modulation instability: The beginning},
  Physica D Nonlinear Phenomena 238 (2009) 540--548.
\newblock \href {http://dx.doi.org/10.1016/j.physd.2008.12.002}
  {\path{doi:10.1016/j.physd.2008.12.002}}.

\bibitem{2011PhRvA..83a3830B}
P.~{B{\'e}jot}, B.~{Kibler}, E.~{Hertz}, B.~{Lavorel}, O.~{Faucher}, {General
  approach to spatiotemporal modulational instability processes}, Physical
  Reviev A 83~(1) (2011) 013830.
\newblock \href {http://dx.doi.org/10.1103/PhysRevA.83.013830}
  {\path{doi:10.1103/PhysRevA.83.013830}}.

\bibitem{2009PhLA..373..675A}
N.~{Akhmediev}, A.~{Ankiewicz}, M.~{Taki}, {Waves that appear from nowhere and
  disappear without a trace}, Physics Letters A 373 (2009) 675--678.
\newblock \href {http://dx.doi.org/10.1016/j.physleta.2008.12.036}
  {\path{doi:10.1016/j.physleta.2008.12.036}}.

\bibitem{Fabrikant1980}
A.~{Fabrikant}, {On non-linear water waves under light wind and Landau type
  equations near the stability threshold}, Wave Motion 2 (1980) 355--359.

\bibitem{Bridges2007}
T.~{Bridges}, F.~{Dias}, {Enhancement of the Benjamin-Feir instability with
  dissipation}, Physics of Fluids 19 (2007) 104104.

\bibitem{Kharif2010}
C.~{Kharif}, R.~A. {Kraenkel}, M.~A. {Manna}, R.~{Thomas}, {The modulational
  instability in deep water under the action of wind and dissipation}, J. Fluid
  Mech. 664 (2010) 138--149.

\bibitem{blennerhassett1980}
P.~J. {Blennerhassett}, {On the generation of waves by wind}, Philosophical
  Transactions of the Royal Society of London. Series A, Mathematical and
  Physical Sciences 298 (1980) 451--494.

\bibitem{Miles1957}
J.~W. {Miles}, {On the generation of surface waves by shear flows}, J. Fluid
  Mech. 3 (1957) 185--204.

\bibitem{Janssen1991}
P.~A.~E.~M. {Janssen}, {Quasi-linear theory of wind-wave generation applied to
  wave forecasting}, Journal of Physical Oceanography 21 (1991) 1631--1642.

\bibitem{Leblanc2007}
S.~{Leblanc}, {Amplification of nonlinear surface waves by wind}, Physics of
  Fluids 19 (2007) 101705.

\bibitem{1986JFM...162..237B}
L.~F. {Bliven}, N.~E. {Huang}, S.~R. {Long}, {Experimental study of the
  influence of wind on Benjamin-Feir sideband instability}, Journal of Fluid
  Mechanics 162 (1986) 237--260.
\newblock \href {http://dx.doi.org/10.1017/S0022112086002033}
  {\path{doi:10.1017/S0022112086002033}}.

\bibitem{1999JFM...401...55W}
T.~{Waseda}, M.~P. {Tulin}, {Experimental study of the stability of deep-water
  wave trains including wind effects}, Journal of Fluid Mechanics 401 (1999)
  55--84.

\bibitem{2005JFM...539..229S}
H.~{Segur}, D.~{Henderson}, J.~{Carter}, J.~{Hammack}, C.-M. {Li}, D.~{Pheiff},
  K.~{Socha}, {Stabilizing the Benjamin-Feir instability}, Journal of Fluid
  Mechanics 539 (2005) 229--271.
\newblock \href {http://dx.doi.org/10.1017/S002211200500563X}
  {\path{doi:10.1017/S002211200500563X}}.

\bibitem{2013JFM...722....5G}
L.~{Grare}, W.~L. {Peirson}, H.~{Branger}, J.~W. {Walker}, J.-P.
  {Giovanangeli}, V.~{Makin}, {Growth and dissipation of wind-forced,
  deep-water waves}, Journal of Fluid Mechanics 722 (2013) 5--50.
\newblock \href {http://dx.doi.org/10.1017/jfm.2013.88}
  {\path{doi:10.1017/jfm.2013.88}}.

\bibitem{2013PhFl...25j1704C}
A.~{Chabchoub}, N.~{Hoffmann}, H.~{Branger}, C.~{Kharif}, N.~{Akhmediev},
  {Experiments on wind-perturbed rogue wave hydrodynamics using the Peregrine
  breather model}, Physics of Fluids 25~(10) (2013) 101704.
\newblock \href {http://arxiv.org/abs/1306.6471} {\path{arXiv:1306.6471}},
  \href {http://dx.doi.org/10.1063/1.4824706} {\path{doi:10.1063/1.4824706}}.

\bibitem{1983Donelan}
M.~A. {Donelan}, {Attenuation of laboratory swell in an adverse wind}, Canada
  Centre of Inland Waters (1983) 11p.

\bibitem{1985JFM...151..427Y}
I.~R. {Young}, R.~J. {Sobey}, {Measurements of the wind-wave energy flux in an
  opposing wind}, Journal of Fluid Mechanics 151 (1985) 427--442.
\newblock \href {http://dx.doi.org/10.1017/S0022112085001033}
  {\path{doi:10.1017/S0022112085001033}}.

\bibitem{2003JFM...487..345P}
W.~L. {Peirson}, A.~W. {Garcia}, S.~E. {Pells}, {Water wave attenuation due to
  opposing wind}, Journal of Fluid Mechanics 487 (2003) 345--365.
\newblock \href {http://dx.doi.org/10.1017/S0022112003004750}
  {\path{doi:10.1017/S0022112003004750}}.

\bibitem{2008Dias}
F.~{Dias}, A.~I. {Dyachenko}, V.~E. {Zakharov}, {Theory of weakly damped
  free-surface flows: a new formulation based on potential flow solutions},
  Physics Letters A 372 (2008) 1297--1302.

\bibitem{OnoratoProment2012}
M.~{Onorato}, D.~{Proment}, {Approximate rogue wave solutions of the forced and
  damped nonlinear Schr\"odinger equation for water waves}, Physics Letters A
  376 (2012) 3057--3059.

\bibitem{2014PhLA..378.1025B}
M.~{Brunetti}, N.~{Marchiando}, N.~{Berti}, J.~{Kasparian}, {Nonlinear fast
  growth of water waves under wind forcing}, Physics Letters A 378 (2014)
  1025--1030.
\newblock \href {http://arxiv.org/abs/1402.1510} {\path{arXiv:1402.1510}},
  \href {http://dx.doi.org/10.1016/j.physleta.2014.02.004}
  {\path{doi:10.1016/j.physleta.2014.02.004}}.

\bibitem{JanssenBook}
P.~{Janssen}, {The interaction of ocean waves and wind}, Cambridge University
  Press, 2009.

\end{thebibliography}

\end{document}